# Probing the tunable multi-cone bandstructure in Bernal bilayer graphene

Anna M. Seiler[1], Nils Jacobsen[1], Martin Statz[1], Noelia Fernandez[1], Francesca Falorsi[1], Kenji Watanabe[2], Takashi Taniguchi[3], Zhiyu Dong[4], Leonid S. Levitov[4], R. Thomas Weitz[1,5]*

[1] 1st Physical Institute, Faculty of Physics, University of Göttingen, Friedrich-Hund-Platz 1, Göttingen 37077, Germany

[2] Research Center for Electronic and Optical Materials, National Institute for Materials Science, 1-1 Namiki, Tsukuba 305-0044, Japan

[3] Research Center for Materials Nanoarchitectonics, National Institute for Materials Science, 1-1 Namiki, Tsukuba 305-0044, Japan

[4] Department of Physics, Massachusetts Institute of Technology, Cambridge, Massachusetts 02139, USA

[5] International Center for Advanced Studies of Energy Conversion (ICASEC), University of Göttingen, Göttingen, Germany

*Corresponding author. Email: thomas.weitz@uni-goettingen.de

**Controlling the bandstructure of Dirac materials is of wide interest in current research but has remained an outstanding challenge for systems such as monolayer graphene. In contrast, Bernal bilayer graphene (BLG) offers a highly flexible platform for tuning the bandstructure, featuring two distinct regimes. In one regime, which is well established and widely used, a tunable bandgap is induced by a large enough transverse displacement field. Another is a gapless metallic band occurring near charge neutrality and at not too strong fields, featuring rich "fine structure" consisting of four linearly-dispersing Dirac cones with opposite chiralities in each valley and van Hove singularities. Even though BLG was extensively studied experimentally in the last two decades, the evidence of this exotic bandstructure is still elusive, likely due to insufficient energy resolution. Here, rather than probing the bandstructure by direct spectroscopy, we use Landau levels as markers of the energy dispersion and carefully analyze the Landau level spectrum in a regime where the cyclotron orbits of electrons or holes in momentum space are small enough to resolve the distinct mini Dirac cones. We identify the presence of four distinct Dirac**

**cones and map out complex topological transitions induced by electric displacement field. These findings introduce a valuable addition to the toolkit for graphene electronics.**

Graphene, a single layer of carbon atoms arranged in a hexagonal lattice, exhibits intriguing electronic properties due to its linearly dispersing bands forming Dirac cones at the K and K' points. Yet, one of the key limitations of graphene is its zero bandgap, which makes it not-well suited for digital electronic applications and controlling electronic interactions. Several attempts have been made to artificially open up a bandgap in monolayer graphene including chemical doping[1–3] and strain engineering[4–7]. However, while these methods may allow to create a gap in an otherwise gapless dispersion, they also create disorder in the system. Opening a tunable band gap in pristine monolayer graphene by electrostatic gating presently appears to be out of reach, since it would require electric field control with atomic precision to induce a potential difference between the two sublattices. Research has expanded to other systems such as Dirac semimetals[8] that, however, also do not allow to electrostatically induce a band gap that can be tuned after device fabrication.

The simple Bernal-stacked bilayer graphene (BLG), to the contrary, does allow electrostatic tunability of a band gap and the high-energy parabolic dispersion – as shown by experimental spectroscopy and transport measurements[9–13] as well as theoretical calculations[14,15]. However, perhaps surprisingly, there is no consensus between experiment and theory regarding the low-energy bandstructure of BLG. For example, quantum Hall measurements identified an eightfold degeneracy of the lowest Landau level (LLL), facilitated by a two-fold spin, valley and orbital degeneracy, consistent with a low-energy parabolic dispersion[12,16,17]. While such quantum Hall measurements provide an elegant means to identify band symmetries, they leave several key questions unanswered. First, upon including higher-order hopping terms, at zero magnetic field and low carrier density, one expects a metallic band that remains gapless at not-too-strong displacement fields (see Fig.S1 and accompanying discussion in the Supplemental

Material). This metallic "fine-structure" band features four Dirac cones with different chiralities at each valley and prominent van Hove singularities, which is inconsistent with experimental findings of quantum oscillations. Other outstanding questions pertain to the presence of strong exchange-driven phases in suspended bilayer graphene[12,17–19] which is hard to reconcile with linear dispersion at low energies. Furthermore, signatures of changes in Fermi surface topology due to trigonal warping have been identified in the case of strong displacement fields where sizable bandgaps are opened. Specifically, an unusual ordering of quantum Hall states is observed in the presence of a strong electric displacement field in strongly biased BLG[20–23], Bernal trilayer graphene[24,25], rhombohedral trilayer graphene[26,27] and Bernal tetralayer graphene[28] which is consistent with the theoretical band structure predicting a trigonally-warped low-energy Fermi surface topology.

The aim of the present work is to resolve the puzzle of why the linearly dispersing bands and electric-field-induced changes of topology (Fig. 1a-c) have not been observed experimentally in BLG. Signs of topological changes in the bandstructure involving a van Hove singularity should be detectable e.g. in a quantizing magnetic field in which the presence of four Dirac cones results in an exotic sequence of Landau levels distinct from the previously studied instances of Landau levels. Accessing this complex field-tunable bandstructure with multiple Dirac cones and field-tunable van Hove singularities is of clear interest for the physics of strongly correlated systems as well as graphene band engineering.

To reveal the detailed low-energy bandstructure and detect electric-field controllable linear dispersing bands, here we have chosen to work with the hexagonal Boron Nitride (hBN) encapsulated Bernal-stacked bilayer graphene sample system over suspended BLG – even though both systems yield samples of similar quality. In the suspended BLG, however, due to the low-dielectric constant of the dielectric (vacuum), the exchange energy scale seems to dominate the low-energy physics leading to a variety of nontrivial groundstates[12,17–19]. Our encapsulated BLG is equipped with graphite top-and bottom gates and two terminal graphite contacts (see Supplemental Material). All measurements were recorded in a cryostat at a temperature of 10 mK employing standard lock-in techniques at 78 Hz and an ac bias current of 1 nA. By varying both gate voltages, we were able to tune the charge carrier density ($n$) and the electric displacement field ($D$) independently. Since we on the one side use two-terminal measurements (i.e. contact resistance cannot be easily determined), and on the other side show measurements down to B=0 the respective LL are not fully developed (quantized) yet, and we hence show d$G$/d$n$ values. We complement our measurements with tight binding bandstructure calculations of expected quantum Hall states in BLG at zero and low $D$-fields, in which we include also the weaker inter-layer coupling parameters $\gamma_3$ and $\gamma_4$ as well as an energy difference between dimer and non-dimer atomic sites $\Delta'$ [14,29] (Fig. 1a) (see Supplemental Material for technical details of the calculations).

We first discuss the tight binding calculations of LL that are used as a tool to later experimentally identify the transition from a metallic band with four Dirac cones to a gapped parabolic dispersion. Without the presence of an interlayer potential difference $U$ and in the case trigonal warping is ignored or irrelevant due to disorder, the bandstructure of BLG exhibits a nearly low-energy parabolic dispersion (Fig 1b (left)). [12,14,30]. Consistent with previous experiments conducted in quantizing magnetic fields $B > 0.5$ T [12,16,18], this leads to an eight-fold degeneracy of the lowest Landau level (LLL) due to spin, valley and orbital degrees of freedom, and to a four-fold degeneracy of all higher Landau levels due to spin and valley degrees of freedom (see Supplemental Material). In case that the low-energy bandstructure at the charge neutrality point can be resolved below a Fermi energy $E \sim \pm 1$ meV, the bandstructure dramatically changes when including trigonal warping, and three off-center and one center cones emerge in each valley (also referred to as mini Dirac cones), resulting from the weaker skew interlayer hopping term $\gamma_3$. The four cones with a Dirac-like spectrum resemble a four-fold copy of the spectrum of monolayer graphene, for which the LLL is shared equally by electrons and holes, overall leading to a 16-fold degeneracy (2 spins, 2 valleys, 4 mini Dirac cones). This would result in the appearance of quantum Hall states with filling factors $\nu = \frac{hn}{eB} = \pm 8$ [16,30,31]. In addition, the second skew interlayer hopping term $\gamma_4$ and on-site parameter $\Delta'$, describing the energy difference between atoms A and A' or B and B', create an energetic asymmetry between these cones [14]. While the center cone of the conduction and valence band meet at zero energy, the off-center cones meet at higher energies (Fig. 1b, c; more information on the impact of $γ_3$, $γ_4$, and $Δ'$ is given in Fig. S2 in the Supplemental Material). In quantum Hall measurements, these changes in the bandstructure can be discerned only at $B$ < 0.2 T since here, the inverse of the magnetic length $l_B = \sqrt{\hbar/(eB)}$ with $\hbar = h/2\pi$ and Planck's constant $h$ is smaller than the distance in momentum space between two adjacent mini Dirac cones (i.e. below the fields at which magnetic breakdown appears) [22,23].

Fig. 1e shows the calculated inverse compressibility ($\partial\mu/\partial n$ with chemical potential $\mu$, $n$ charge carrier density) as function of $n$ and $B$ at $D$ = 0 V/nm (Landau fan diagram). Here, larger energy gaps in the Landau level spectrum (Fig. 1f) manifest as prominent peaks corresponding to quantum Hall states that are labeled by numerals (the calculations include $γ_0$, $γ_1$, $γ_3$, $γ_4$ and $Δ'$, spin splitting is included manually in the LL for the figures, both valleys are included in the calculation and are fully degenerate; see Supplemental Material for further details). While quantum Hall states with $ν$ = ±8 indeed exhibit the largest compressibility and go down to the lowest $B$ in the valence/ conduction band also the quantum Hall state with $ν$ = -4 (but not $ν$ = +4) is very robust, i.e., it can be resolved until very low $B$ (Fig. 1e), which is a manifestation of the electron-hole asymmetry. Neglecting the spin and valley degrees of freedom, the three off-center cones exhibit a three-fold degenerate LLL and are shifted to higher energies. Thus, the center cone LLL is non-degenerate with the LLL that belong to the off-center cones (Fig 1f). Since the LLL is shared between electrons and holes, the non-degenerate center cone as well as one of three LL originating from the three off-center cones contribute to hole transport and give rise to quantum Hall states with $ν$ = -8 and $ν$ = -4 respectively. The other two LL stemming from the three off-center cones contribute to electron transport and give rise to a quantum Hall state with $ν$ = +8. The quantum Hall state with $ν$ = +4 only emerges at larger $B$ where the degenerate LL diverge. With increasing $n$, the conventional sequence quantum Hall states with filling factors $\nu$ = ±12, ±16, ±20 is recovered and the Fermi level lies above the Lifshitz transition where the Fermi surface is fully connected.

The theory thus shows in great detail how the presence of four Dirac cones can unambiguously be identified in experiment. Fig. 1g shows the normalized derivative of the measured two-terminal conductance |dG/dn| as a function of $n$ and $B$ at $D$ = 0 V/nm. Quantum Hall states appear as plateaus in the conductance and thus as dips in |dG/dn| and can be assigned by their corresponding slopes in the Landau fan diagram (see Supplemental Material). Consistent with our theoretical simulations, quantum Hall states with $\nu$ = ±8 are the most robust and can be observed at the smallest $B$, down to $B$ ≈ 0.05 T which reveals the presence of four mini Dirac cones. Additionally, due to electron-hole asymmetry, the quantum Hall state with $\nu$ = -4 appears at slightly larger magnetic fields (B ≈ 0.15 T), while the $\nu$ = +4 quantum Hall state only appears above 0.2 T when the magnetic breakdown occurs (indicated by dashed lines in Fig 1e and g). At magnetic fields above 0.3 T, a sequence of even integer quantum Hall states appears which is consistent with previous measurements in freestanding BLG[12,17,18,32] and which reveals the high quality of our sample. Here, the spin degeneracy is likely lifted due to Coulomb interactions resulting in a two-fold degeneracy (valley) instead of the predicted four-fold degeneracy (spin and valley)[30,33,34]. Notably, some of the non-four-fold degenerate quantum Hall states including the quantum Hall states with v = ±6 also go down to below 0.3 T and then merge with the quantum Hall states with v = ±4 and demand further investigation. Since spin and valley splitting are both neglected in our theoretical simulations, this two-fold degeneracy is only observed in our experimental data but not visible in the calculated inverse compressibility.

While the measurements at zero displacement fields show the existence of four Dirac cones near charge neutrality, we now show the tunability of the Dirac cones followed by bandgap opening controlled by $D$. We note in passing that for very small D-fields while the four Dirac cones will be individually gapped, there will be no overall gap in the spectrum due to the inherent energetic offset between the center and the three off-center Dirac cones. While theoretical tight binding calculations of this regime are discussed in the Supplemental Material and are shown in Fig. S1, this regime is not within our experimental resolution. We first discuss our calculations and corresponding charge transport measurements at constant small $D$, where already an overall bandgap has opened up in the previously linear Dirac spectrum of BLG. This goes along with drastic changes of the bandstructure: the center cone diminishes whereas the three off-center cones (Fig. 2a,b) with however a parabolic dispersion remain, which we consequently refer to as pockets.

In the valence band, where the three off-center cones already dominate at $D = 0$ due to electron-hole asymmetry, the number of pockets changes from four to three at finite $D$ (Fig. 2b) resulting in a reordering of expected quantum Hall states (Fig. 2c-e)[29,35]. We expect the LLL to be six-fold degenerate at low $B$ due to the remaining three leg pockets (3 pockets, 2 spins, the 2 valleys are degenerate, Fig. 2d,e)[22,23] and the $\nu = -6$ quantum Hall state is expected to be the most robust for hole doping (Fig. 2c). These theoretical considerations are confirmed by our measurements. As shown in Fig. 2e, for $D = 50$ mV/nm the $\nu = -6$ quantum Hall state can be resolved down to very low magnetic fields of $B = 100$ mT. Surprisingly, this also holds for the $\nu = -3$ quantum Hall state which could result from spin or valley polarization at low $B$ and low $n$ due to Stoner ferromagnetism that can occur in the vicinity of the Lifshitz transition [20,21,36]. At $B = 600$ mT, a sudden change in the degeneracy of Landau level takes place for $n < 0$ which can be attributed to the magnetic breakdown. Here the effects of the trigonal warping are no longer relevant and we can observe all integer quantum Hall states. It is worth noting that at larger densities, quantum Hall states with $\nu = -7$ and $\nu = -9$ appear below $B = 500$ mT. In this regime, the Fermi energy level lies above the Lifshitz transition and quantum Hall states start to become valley and spin polarized with increasing magnetic field.

The effects of band flattening and disappearing of the low-energy Dirac spectrum can be also seen in the conduction band (Fig. 2a) where the center pocket also becomes less prominent with increasing $D$. However, due to electron-hole asymmetry the center cone is still dominating at $U = 17$ meV and the band becomes flatter with increasing $U$ until $U \approx 60$ meV[37]. At $U = 17$ meV, the degeneracy of quantum Hall states is not as much affected by trigonal warping as in the valence band and quantum Hall states with even ν appear first in the magnetic field in our conductance measurements (Fig. 2e). Remarkably, the quantum Hall states with ν = +3 and ν = +4 disappear at a magnetic field of 0.5 T and then reemerge at about 0.6 T (Fig. 2c,e) resulting from a crossing of two bands that correspond to different valleys (Fig. 2d).

The active control and lifting of the four-fold Dirac spectrum can be also traced by controlling the displacement field at constant B (Fig. 3a,b). For example, at $B = 0.25$ T (Fig. 3a) the magnetic breakdown has already occurred for low $D$ resulting in the appearance of eight-fold degenerate quantum Hall states at $D = 0$ mV/nm (2 valleys, 2 spins, 2 orbits). At 3 mV/nm < $|D|$ < 25 mV/nm the quantum Hall state with ν = ±2 appears due to valley polarization[18] while at $|D| > 20$ mV/nm the three pockets can be resolved individually and a crossover from a two-fold (two spins) to a three-fold degenerate Landau level spectrum (three pockets) appears at hole doping (yellow circle in Fig. 3a). Higher LL are four-fold degenerate (2 valleys, 2 spins) above a Lifshitz transition since their corresponding Fermi surface is fully connected. For larger $B$, e.g. at $B = 0.4$ T (Fig. 3b), the crossover to the parabolic shifts to larger $D$ where the three pockets are more pronounced. Here also the crossings of LL stemming from different valleys in the conduction band can be discerned in the $n$ vs. $B$ plot.

In conclusion, we present measurements demonstrating, for the first time, that bilayer graphene exhibits a highly tunable bandstructure at low energies where four distinct Dirac cones, under tuning transverse field, undergo topological transitions and merge into a parabolic band or into three pockets with a gapped parabolic dispersion. The topological transitions result in a complex series of Landau levels that we extract by virtue of numerical diagonalization methods based on a realistic tight binding model and measurements in high quality hBN-encapsulated samples. These results show that the simple and seemingly well understood Bernal bilayer graphene is a true example of a long sought tunable Dirac material with linear dispersion at low energies. This, along with tunable van Hove singularities, makes it a promising material for exploring novel ordered states of interacting electrons and developing low-energy fast electronics. The role of electron interactions and trigonal warping effects, in particular their impact on the renormalized many-body bandstructure, is an interesting topic for future work that can be addressed in high quality freestanding BLG samples where interaction effects dominate even at low electric fields[12,18].

**Acknowledgements**

R.T.W. and A.M.S. acknowledge funding from the Deutsche Forschungsgemeinschaft (DFG, German Research Foundation) under the SFB 1073 project B10. N.J. acknowledges funding from the International Center for Advanced Studies of Energy Conversion (ICASEC). R.T.W. acknowledges funding from the DFG SPP 2244. K.W. and T.T. acknowledge support from the JSPS KAKENHI (Grant Numbers 21H05233 and 23H02052) and World Premier International Research Center Initiative (WPI), MEXT, Japan.

**Author contributions**

A.M.S. and M.S. fabricated the devices and conducted the measurements. A.M.S. conducted the data analysis. N.J. performed the theoretical simulations supervised by L.S.L.. K.W. and T.T. grew the hexagonal nitride crystals. All authors discussed and interpreted the data. R.T.W. supervised the experiments and the analysis. The manuscript was prepared by A.M.S., N.J., L.S.L. and R.T.W. with input from all authors.

**Corresponding authors**

R. Thomas Weitz (thomas.weitz@uni-goettingen.de)

**Competing interests**

Authors declare no competing interests.

**Data availability**

The datasets generated during and/or analyzed during the current study are available from the corresponding author on reasonable request.

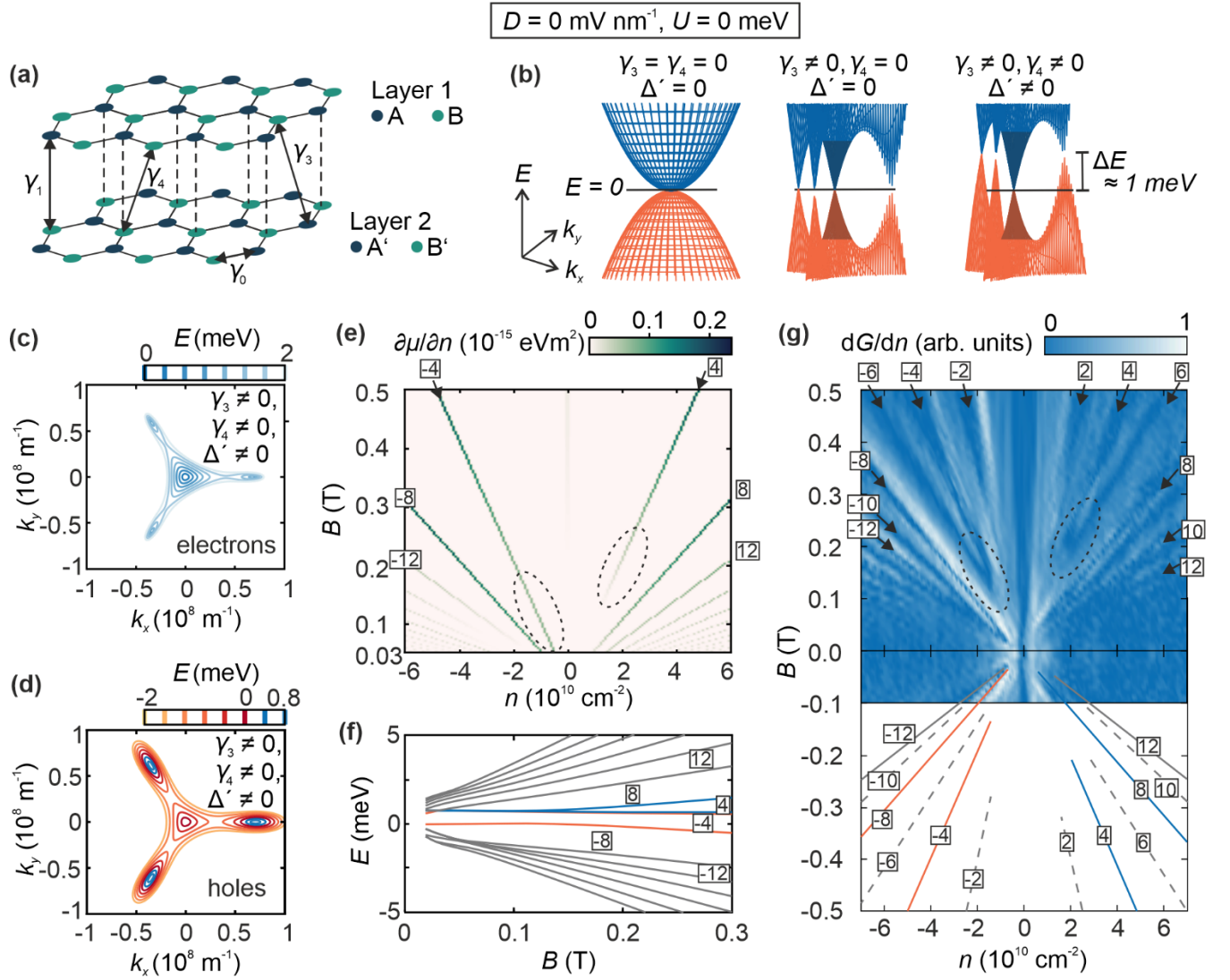

FIG. 1. Lattice, bandstructure and quantum Hall states of Bernal bilayer graphene at $D$ = 0 mV/nm. **(a)** Lattice structure and hopping parameters of Bernal bilayer graphene. **(b)** Bandstructure of bilayer graphene in an energy range of -2 mV to +2 mV at $D$ = 0 mV/nm calculated with a tight binding model including various subsets of coupling parameters, featuring four Dirac cones of different chiralities and three van Hove singularities in each valley. The center cones are shaded darker. **(c,d)** Fermi surface contour of the conduction band **(c)** and valence band **(d)** of bilayer graphene at different Fermi energy levels. **(e)** Calculated inverse compressibility ($\partial\mu/\partial n$) as a function of charge carrier density and magnetic field at $D$ = 0 mV/nm and temperature $T$ = 0.1 K. The corresponding quantum Hall states are labeled by numerals. The regions in which quantum Hall states with filling factors $\nu = \pm 4$ terminate are highlighted by dashed circles. **(f)** Evolution of Landau levels as a function of magnetic field at $D$ = 0 mV/nm. The four lowest Landau levels are colored, whereas Landau levels contributing to hole transport are colored in red and Landau levels contributing to electron transport are colored in blue. The lowest red colored Landau level is coming from the center mini Dirac cone, the other three lowest Landau levels are coming from the three off-center mini Dirac cones. Filling factors are indicated by numerals. A larger version of this plot is shown in Fig. S2. **(g)** Derivative of the normalized conductance measured as a function of the charge carrier

density and the magnetic field at $D$ = 0 mV/nm. Blue regions correspond to vanishing differential conductance (i.e. a conductance plateau). The slopes of the lowest quantum Hall states are traced by lines in the mirror image. Solid lines correspond to the non-interaction induced quantum Hall states that allow for comparison with theory (Fig. 1e). Their colors are adapted from Fig. 1f. Dashed lines correspond to interaction-induced quantum Hall states in which the spin or valley degrees of freedom is broken. The regions in which quantum Hall states with filling factors ν = ±4 end are highlighted by dashed circles.

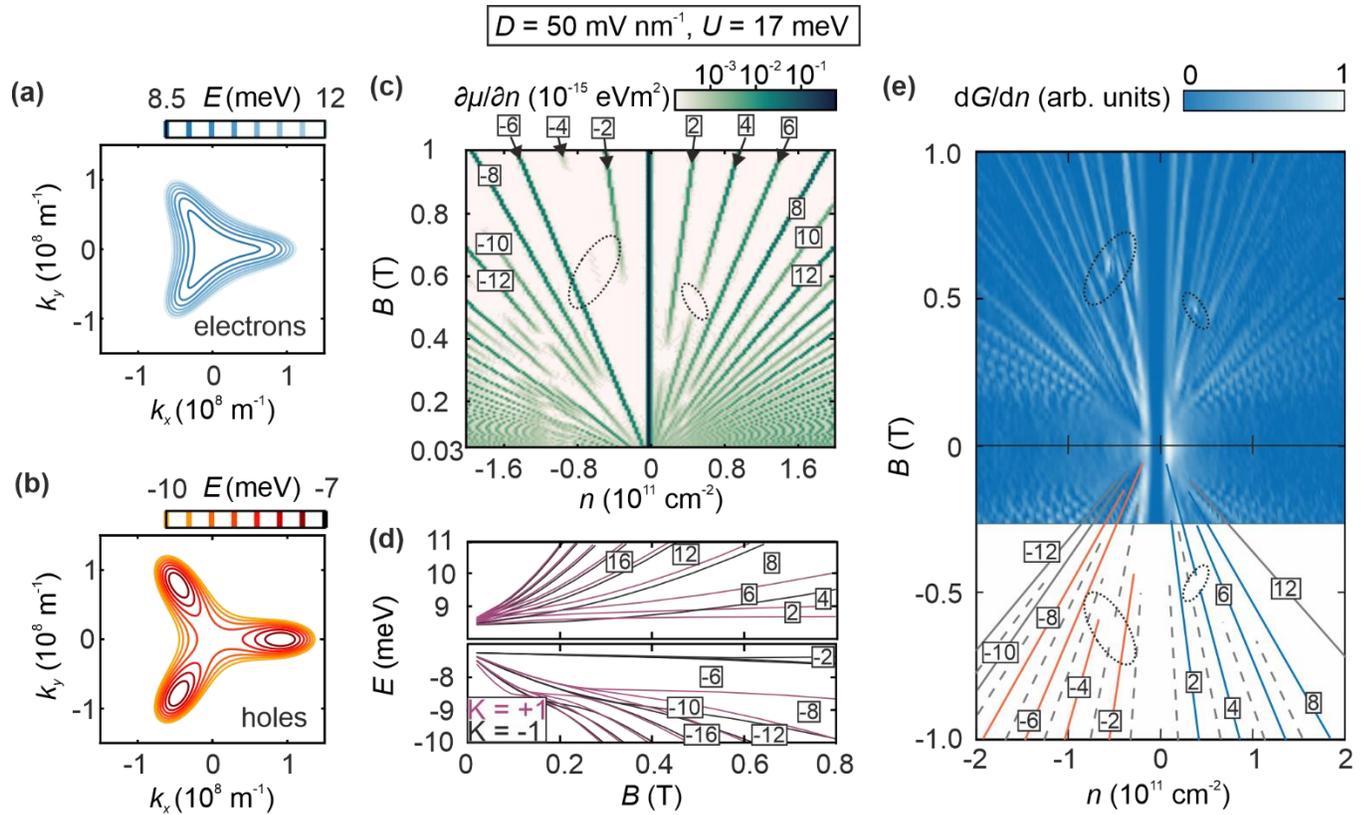

FIG. 2. Bandstructure and quantum Hall states of bilayer graphene at finite $D$. **(a,b)** Fermi surface contour of the conduction band **(a)** and valence band **(b)** of bilayer graphene at different Fermi energy levels at $U = 17$ meV. **(c)** Calculated inverse compressibility $(\partial\mu/\partial n)$ as a function of charge carrier density and magnetic field at $U = 17$ meV and temperature $T = 0.1$ K. The corresponding quantum Hall states are labeled by numerals. Regions corresponding to Landau level crossings are marked by dotted circles. **(d)** Evolution of Landau levels as a function of the magnetic field. **(e)** Derivative of the normalized conductance measured as a function of the charge carrier density and the magnetic field at $D = 50$ mV/nm. Solid lines correspond to the non-interaction induced quantum Hall states that allow for comparison with theory (Fig. 2c). Dashed lines correspond to interaction-induced quantum Hall states in which the spin degree of freedom is broken. The corresponding quantum Hall states are labeled by numerals. The regions corresponding to Landau level crossings are marked by dotted circles.

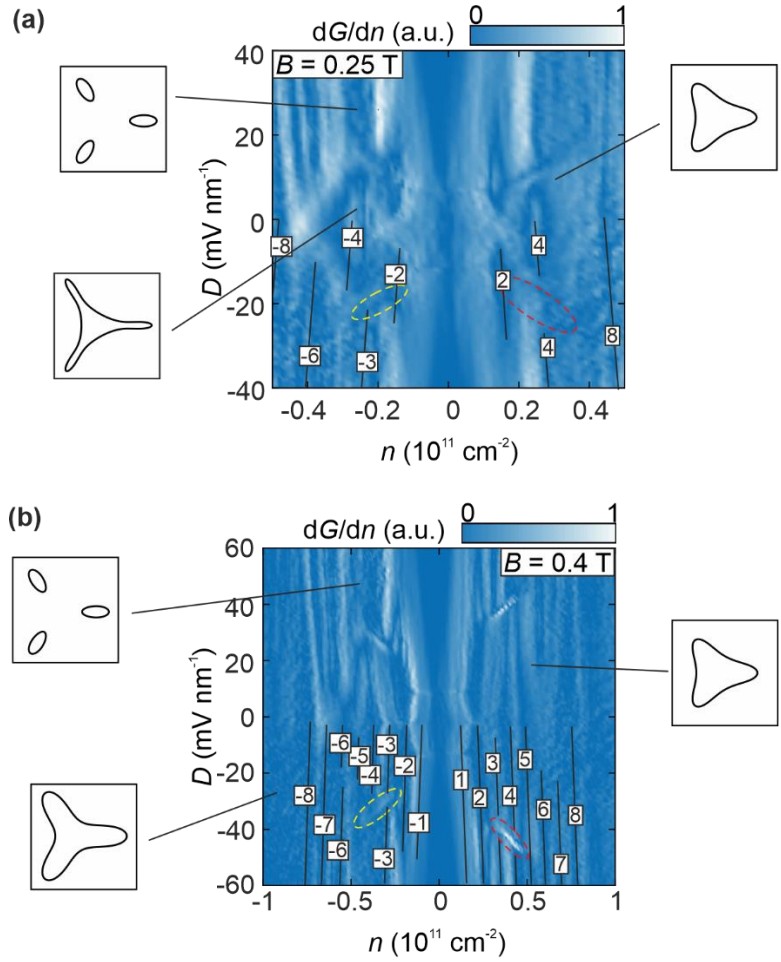

FIG 3. Changing the degeneracy of quantum Hall states due to magnetic breakdown induced by an electric displacement field. **(a, b)** Derivative of the normalized conductance measured as a function of the charge carrier density and electric displacement field at $B$ = 0.25 T **(a)** and $B$ = 0.4 T **(b)**. Quantum Hall states are labeled by numerals and traced by lines. Note, that the quantum Hall states are symmetric for positive and negative values of $D$ but the labelling was restricted to $D < 0$ for better visibility. Transitions between quantum Hall states due to trigonal warping and electron-hole asymmetry are highlighted by yellow dotted lines, crossings between Landau levels of different K valleys are highlighted by red dotted circles. Quantum Hall states are three-fold degenerate at large $D$ where the pockets are more pronounced and the magnetic breakdown has not yet accord. Schematics of Fermi contours corresponding to regions which different Fermi surface topologies are shown in the insets. Apart from the number of Fermi surfaces, the degeneracy of quantum Hall states is also affected by spin and valley polarization at large $B$ and $D$.

# Supplementary Material

## for

## Probing the tunable multi-cone bandstructure in Bernal bilayer graphene

In this supplementary material we give more details on the tight-binding and inverse compressibility calculations described in the main text. Furthermore, we give a characterization of the device discussed in the main text and present data conducted in a second device.

### A. Tight-binding calculations

The Landau level calculations were based on a realistic tight binding model for the $\pi$-electrons as described in Ref. [14].

The model includes different hopping processes which are described by different parameters: $\gamma_0$ describes the tunneling for neighboring sites within a single graphene sheet; $\gamma_1$ accounts for hopping processes between the aligned lattice sites of the two sheets, the so-called dimer sites. Finally, $\gamma_3$ and $\gamma_4$ are the hopping parameters between non-dimer sites; $\gamma_3$ causes the trigonal warping of the Fermi sea at low carrier densities. Further parameters are $\Delta'$, an energy difference between dimer and non-dimer sites, and the interlayer potential $U$ accounting for an out-of-plane displacement field. The magnitudes of these parameters were taken from Ref. [29].

We used a two-band model which is reduced to the non-dimer sites and includes direct hoppings via $\gamma_3$ and $\gamma_4$, as well as hoppings via dimer sites. It is expected to capture the correct low-energy physics if $\gamma_0$ and $\gamma_1$ are the relevant energy scales of the problems.

Writing $\pi = \xi p_x + i p_y$ and $\pi^+ = \xi p_x - i p_y$, where $\xi = +1, -1$ denotes the valley index and $p_x$ and $p_y$ are the x and y components of the momentum vector, one may decompose the Hamiltonian as $h = h_0 + h_w + h_{as} + h_U$ with

$$h_0 = -\frac{1}{2m}\begin{pmatrix} 0 & (\pi^+)^2 \\ \pi^2 & 0 \end{pmatrix},$$

$$h_w = v_3 \begin{pmatrix} 0 & \pi \\ \pi^+ & 0 \end{pmatrix} - \frac{v_3 a}{4\sqrt{3}\hbar}\begin{pmatrix} 0 & (\pi^+)^2 \\ \pi^2 & 0 \end{pmatrix},$$

$$h_{as} = \left(\frac{2vv_4}{\gamma_1} + \frac{\Delta' v^2}{\gamma_1^2}\right)\begin{pmatrix} \pi^+\pi & 0 \\ 0 & \pi\pi^+ \end{pmatrix},$$

$$h_U = -\frac{U}{2}\left[\begin{pmatrix} 1 & 0 \\ 0 & -1 \end{pmatrix} - \frac{2v^2}{\gamma_1^2}\begin{pmatrix} \pi^+\pi & 0 \\ 0 & -\pi\pi^+ \end{pmatrix}\right],$$

where we have introduced the velocities $v = \frac{\sqrt{3}a\gamma_0}{2\hbar}$, $v_3 = \frac{\sqrt{3}a\gamma_3}{2\hbar}$, $v_4 = \frac{\sqrt{3}a\gamma_4}{2\hbar}$ with the lattice constant $a$ and the band mass parameter $m = \frac{\gamma_1}{2v^2}$. A diagonalization yields momentum space representations of two component wave functions $\Psi = \frac{1}{\sqrt{2}}(\psi_1, \psi_2)$, with the components $\psi_{1(2)}$ specifying the wave functions projections on the non-dimer sites.

The first term ($h_0$) describes a simple parabolic two-band model with the dominant hopping processes between non-dimer sites, the second ($h_w$) accounts for trigonal warping, the third one ($h_{as}$) introduces an intrinsic electron-hole asymmetry and the fourth one ($h_U$) describes coupling to external fields.

The effect of the latter may be understood by realizing that a displacement field results in a potential difference $U$ between the two graphene sheets. A rigorous estimation of the magnitude of the potential requires a self-consistent computation that includes the screening effects due to the redistribution of charge carriers between the two sheets in the presence of a displacement field [14].

Without aiming for an exact quantitative description, we estimate the rough magnitude of $U$ via a simple plate-capacitor calculation as $U = ecD$ where $c = 3.35$ Å is the interlayer spacing. This estimate assumes minor importance of screening effects at small displacement fields.

The out-of-plane magnetic field of magnitude $B$ is introduced by adding a contribution from the vector potential to canonical momenta. In the Landau gauge this modifies the momentum operator according to $\pi = -i\xi\hbar\partial_x + \hbar\partial_y - ieBx$ and $\pi^+ = -i\xi\hbar\partial_x - \hbar\partial_y + ieBx$. The operators satisfy the same algebraic relation as the ladder operators of the harmonic oscillator, namely $[\pi, \pi^+] = -2\hbar Be\xi$. One may use this fact to define the operators $a = -\frac{i}{\sqrt{2\hbar eB}}\pi$ and $a^+ = \frac{i}{\sqrt{2\hbar eB}}\pi^+$ for the $K_+$-valley or $a^+ = -\frac{i}{\sqrt{2\hbar eB}}\pi$ and $a = -\frac{i}{\sqrt{2\hbar eB}}\pi^+$ for the $K_-$-valley. These operators act like raising and lowering operators for an oscillator with the cyclotron frequency $\omega_c = \frac{eB}{m}$. Their action on the oscillator wave functions reads $a\Phi_l = \sqrt{l}\Phi_{l-1}$ and $a^+\Phi_l = \sqrt{l+1}\Phi_{l+1}$

The Hamiltonian in the $K_+$-valley may then be rewritten as

$$\hat{h} = \begin{bmatrix} -\frac{U}{2} + C_+ a^+ a & A(a^+)^2 - iRa \\ Aa^2 + iRa^+ & \frac{U}{2} + C_- aa^+ \end{bmatrix}$$

with

$$A = \hbar\omega_c \left(1 + \frac{\gamma_1\gamma_3}{6\gamma_0^2}\right),$$

$$R = \frac{\gamma_3}{\gamma_0}\sqrt{\gamma_1\hbar\omega_c}$$

and

$$C_\pm = \hbar\omega_c \left(\frac{2\gamma_4}{\gamma_0} + \frac{\Delta' \pm U}{\gamma_1}\right).$$

In the opposite valley the roles of creators and annihilators are interchanged.

In the presence of trigonal warping interaction the eigenvalue problem has no analytic solution. In order to get a numerical solution, we used a matrix representation of the Hamiltonian in a truncated basis. This method is expected to give good results for the low-energy spectrum, as the discarded high-energy states do not hybridize with those at low energy.

In the $K_+$-valley, using

$$\left\{ \begin{bmatrix} \phi_0 \\ 0 \end{bmatrix}, \begin{bmatrix} \phi_1 \\ 0 \end{bmatrix}, \frac{1}{\sqrt{2}} \begin{bmatrix} \phi_n \\ \sigma\phi_{n-2} \end{bmatrix} \middle| (n \geq 2) \; \sigma = \pm \right\},$$

the matrix elements evaluate to

$$\langle 0|h|0\rangle = \frac{-U}{2} \qquad \langle 3,\sigma|h|0\rangle = \frac{\sigma iR}{\sqrt{2}}$$

$$\langle 1|h|1\rangle = \frac{-U}{2} + C_+ \qquad \langle 4,\sigma|h|1\rangle = \sigma iR$$

$$\langle n,\sigma|h|n,\sigma'\rangle = \sigma\delta_{\sigma\sigma'}A\sqrt{n(n-1)} - \frac{U}{2}(1-\delta_{\sigma\sigma'}) + \frac{1}{2}(C_+ n + \sigma\sigma' C_-(n-1))$$

$$\langle n+3,\sigma|h|n,\sigma'\rangle = \frac{\sigma iR\sqrt{n+1}}{2}.$$

All remaining matrix elements follow from the requirement of the matrix being Hermitian.

Analogously, for the $K_-$- valley we used

$$\left\{ \begin{bmatrix} 0 \\ \phi_0 \end{bmatrix}, \begin{bmatrix} 0 \\ \phi_1 \end{bmatrix}, \frac{1}{\sqrt{2}} \begin{bmatrix} \phi_{n-2} \\ \sigma\phi_n \end{bmatrix} \middle| (n \geq 2) \; \sigma = \pm \right\}$$

to obtain

$$\langle 0|h|0\rangle = \frac{U}{2} \qquad \langle 3,\sigma|h|0\rangle = \frac{-iR}{\sqrt{2}},$$

$$\langle 1|h|1\rangle = \frac{U}{2} + C_- \qquad \langle 4,\sigma|h|1\rangle = -iR,$$

$$\langle n,\sigma|h|n,\sigma'\rangle = \sigma\delta_{\sigma\sigma'}A\sqrt{n(n-1)} - \frac{U}{2}(1-\delta_{\sigma\sigma'}) + \frac{1}{2}(C_+(n-1) + \sigma\sigma' C_- n),$$

and

$$\langle n+3,\sigma|h|n,\sigma'\rangle = \frac{-\sigma' iR\sqrt{n+1}}{2}.$$

For the calculation, an upper cutoff for the Landau level index was set at $n_{max} = 300$ by observing the convergence behavior of the low-lying energy levels.

To illustrate the impact of $\gamma_3$, $\gamma_4$, and $\Delta'$ on the low-energy band and Landau levels, the band structure is shown in Figs. S2 and S3 with $\gamma_3$, $\gamma_4$, and $\Delta'$ not included in (a), with only $\gamma_3$ included in (b) and $\gamma_3$, $\gamma_4$, and $\Delta'$ all included in (c). Also, Figs. S2 and S3 show the evolution of Landau levels as a function of the magnetic field, as well as the inverse compressibility in the quantized Hall regime that was calculated as a function of charge carrier density and magnetic field.

The electron-hole asymmetry in the bandstructure in combination with trigonal warping gives rise to a semi-metallic behavior at a low interlayer bias $U$. The interlayer bias gaps out the cones individually, however if sufficiently small, the gaps do not exceed the energetic offset of the leg pocket with respect to the center pocket, such that a global gap only emerges at a certain threshold value of U ≈ 1 meV (Fig. S1). Below this interlayer potential, a finite density of states remains in the overlap region of the electron and the hole band. Owing to the small energy window for this phenomenon to appear, this regime is not directly accessible by spectroscopy methods. The magnetotransport measurements presented in the main text supported by the Landau level calculations serve as an indirect proof for the existence of this regime.

### B. Calculation of the inverse compressibility

The inverse compressibility is defined by $\frac{\partial \mu}{\partial n}$. In contrast to transport coefficients such as the conductance $G$, this quantity can be extracted directly from the Landau level spectrum, which makes it a suitable quantity for theoretical considerations. While distinct from conductance, its behavior is related to that of conductance: Divergences in the inverse compressibility indicate positions of filled Landau levels where the conductance exhibits a plateau.

By fixing the temperature $T$, it is straightforward to calculate the charge carrier density $n$ as a function of $\mu$ by populating the energy levels $\{\epsilon_i\}$ according to the Fermi function. Each level comes with a degeneracy of $g = \frac{ABe}{2\pi\hbar}$, where $A$ is system area, leading to a charge carrier density of $n(\mu) = 2\sum_i \frac{Be}{2\pi\hbar} \frac{1}{1+\exp\left(\frac{\epsilon_i - \mu}{k_B T}\right)}$ with the prefactor of 2 accounts for spin degeneracy. The chemical potential $\mu(n)$ and the inverse compressibility were obtained by a numerical inversion of this function. To this end, we defined a reservoir $M$ of $10^5$ equally spaced values of $\mu$ in a range that was roughly adjusted to the lowest and highest Landau level energies accessible the considered charge carrier densities. For these values of $\mu$, the carrier densities were computed. The contribution of the lower half of the spectrum (the hole Landau levels) had to be subtracted as an offset.

Fixing the carrier density to $n^*$, the corresponding chemical potential could then be determined as $\mu(n^*) = \min(\mu \in M \mid n(\mu) > n^*)$. The inverse function defined in this way may attain all values from the reservoir from the bottom to the top when $n$ is increased. The numerical error of this procedure is controlled by the spacing within the reservoir. For the practical implementation the temperature entering in the Fermi distribution was chosen to be 0.1 K. This is higher than the usual cryogenic temperatures of the actual experimental realization. However, the resulting broadening may also mimic the finite width of Landau levels due to disorder in the sample.

## C. Device characterization

We performed quantum Hall measurements in two different devices. Measurements conducted in one device are shown in the main text. This device and its fabrication is described in detail in Ref. [20] where the same device is denoted as Device A. Measurements conducted in a second device are discussed in Section D and Fig. S7.

In our dually gated bilayer graphene samples the charge carrier density $n$ as well as the electric displacement field $D$ can be tuned individually via the use of graphite top and bottom gates. They are defined as

$$n = \varepsilon_0 \, \varepsilon_{hBN} \, (V_t/d_t + V_b/d_b)/e$$

and $\quad D = \varepsilon_{hBN} \, (V_t/d_t - V_b/d_b )/2$

where $V_t$ ($V_b$) is the gate voltage applied to the top (bottom) gate, $d_t$ ($d_b$) the thickness of the upper (lower) hBN flake serving as a dielectric, $e$ the charge of an electron, $\varepsilon_{hBN}$ the dielectric constant of hBN and $\varepsilon_0$ the vacuum permittivity.

In order to determine $\varepsilon_{hBN}$ and to thereby assign $n$ and $D$, integer quantum Hall plateaus at finite magnetic fields were aligned with their corresponding slopes in the fan diagram. All observed LL crossings show excellent agreement with those observed previously (see Ref. [20] where data from the same device is shown). For example, at $B$ = 0.4 T (Fig. 3b) one can see the known LL crossings of the $v$ = ± 1 and $v$ = 0 quantum Hall states at $D ≈ 15$ mV/nm as well as crossings at $v$ = ± 2 and $D ≈ 0$ mV/nm.

Having aligned the sample by using the slopes of the quantum Hall states would in principle allow to determine the contact resistance by comparing the measured resistance with the expected quantum Hall resistance. However, due to the use of graphite contacts in a two-terminal device configuration the contact resistance increases linearly with increasing magnetic field (see for example Fig. S4, more details are given in Ref. [20]). Furthermore, there is a line of decreased conductance across zero displacement field (Fig. S4). This region is only bottom- but not top-gate dependent and stems from the region of the BLG that is located below the graphite contacts where the top graphite contacts screen the field of the top gate but not of the bottom gate. Thus, the contact resistance is additionally dependent on the top gate voltage ( therefore also on $n$ and $D$). To not confuse the reader with the line of decreased conductance we only show the derivative of the conductance in the main text. Another advantage of showing the derivative of the conductance is that it allows to track quantum Hall states at lower magnetic fields where the conductance is not fully quantized yet as traceable fluctuations near incompressible quantum Hall states can appear [18,38,39]. Exemplary, the conductance including a subtracted contact resistance is shown in Fig. S5 for D = 0 mV/nm and in Fig. S6 for D = 50 mV/nm. Here a contact resistance was subtracted that linearly increases with $B$. However, we did not account for the dependence on the charge carrier density. Therefore, the resistance values are only valid in a small density regime (negative densities close to the band edge). In Fig. S5, we included data taken at larger magnetic fields up to $B$ = 1.5 T which we did not show in the main text. At $B$ > 0.6 T and D = 0 V/nm the quantum Hall states are fully polarized due to additional valley imbalances implying a small residual displacement field. In agreement with previous studies [12,18,33,34], the even integer quantum Hall states still show wider plateaus compared to the odd integer quantum Hall states.

### D. Measurements conducted in a second device

Electrical measurements conducted in a second device are shown in Fig. S7. This device was fabricated the same way as the device described in the main text [20] but it exhibits four graphite contacts instead of two. The corresponding measurement configuration is shown in Fig. S7a. The measurements show agreement with the theoretical simulations and the electrical measurements discussed in the main text.

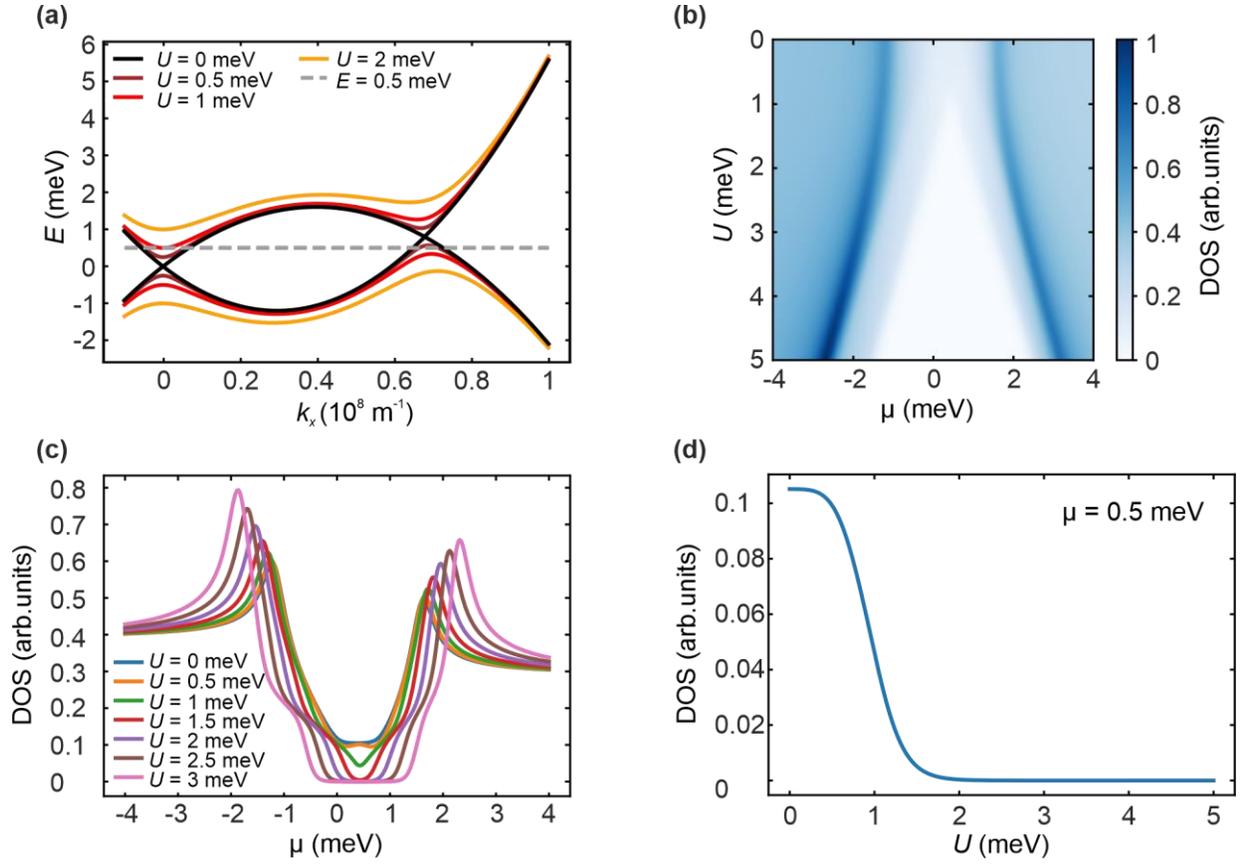

**FIG. S1.** Tight binding calculations revealing a semi-metallic regime near the band edge at small $U$. **(a)** Energy bands as a function of $k_x$. For $U$ = 0.5 meV there is no overall gap in the spectrum even though the single pockets are gapped. **(b)** DOS as a function of $\mu$ and $U$ at $T$ = 1 K. The DOS is approximately constant around of $\mu$ = 0.5 meV for $U$ < 0.5 meV **(c, d)** Linecuts of the DOS at constant $U$ **(c)** and constant $\mu$ **(d)**, respectively.

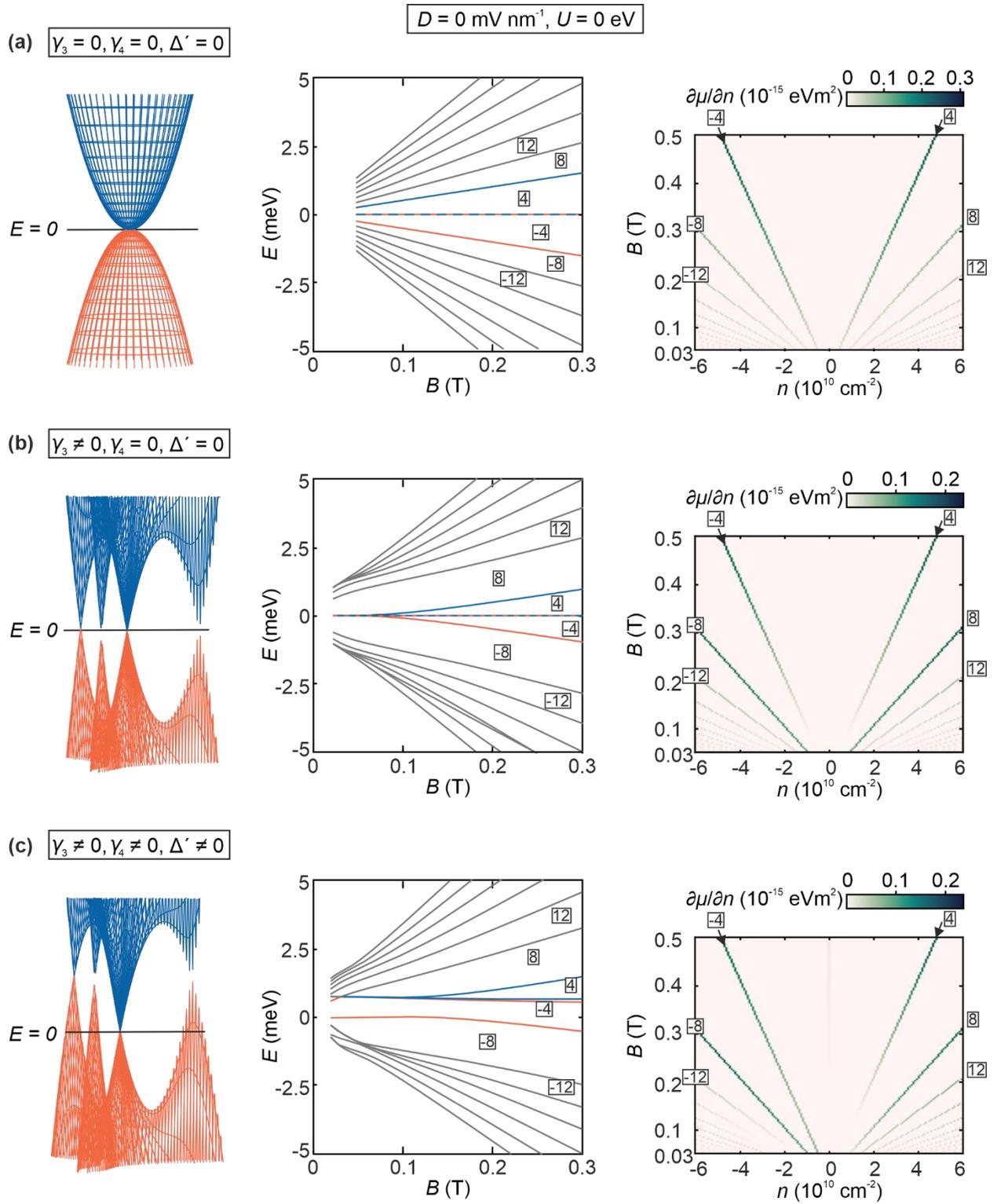

**FIG. S2**. Impact of Coupling parameters $\gamma_3$, $\gamma_4$, and $\Delta'$ on the band structure shown in an energy range from -2 meV to + 2 meV (left panel), the evolution of Landau levels as a function of the magnetic field (middle panel), and the calculated inverse compressibility as a function of charge carrier density and magnetic field (right panel) at *U* = 0. (a) $\gamma_3$, $\gamma_4$, and $\Delta'$ are not included into the calculations. The parabolic band structure results in four-fold degenerate Landau level and an eight-fold degenerate lowest Landau level. This results in the appearance of quantum Hall states with *v* = -12, -8, -4, +4, +8, +12, … at very low magnetic fields. (b) $\gamma_3$ is included into the calculations, $\gamma_4$, and $\Delta'$ are not included. At low energies, the band structure consists of four mini Dirac cones resulting in a 16-fold degenerate lowest Landau level and in the appearance of quantum Hall states with *v* = -12, -8, +8, +12, … at very low magnetic fields. Quantum Hall states with *v* = -4, +4 appear at B > 0.1 T after the magnetic breakdown has occurred. (c) $\gamma_3$, $\gamma_4$, and $\Delta'$ are all included into the calculations. This case is discussed in detail in the main text.

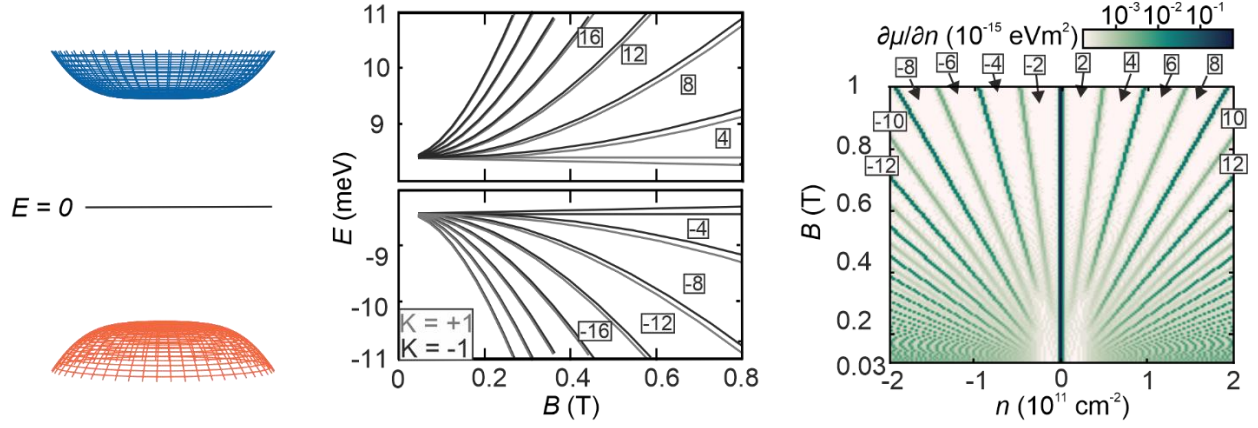
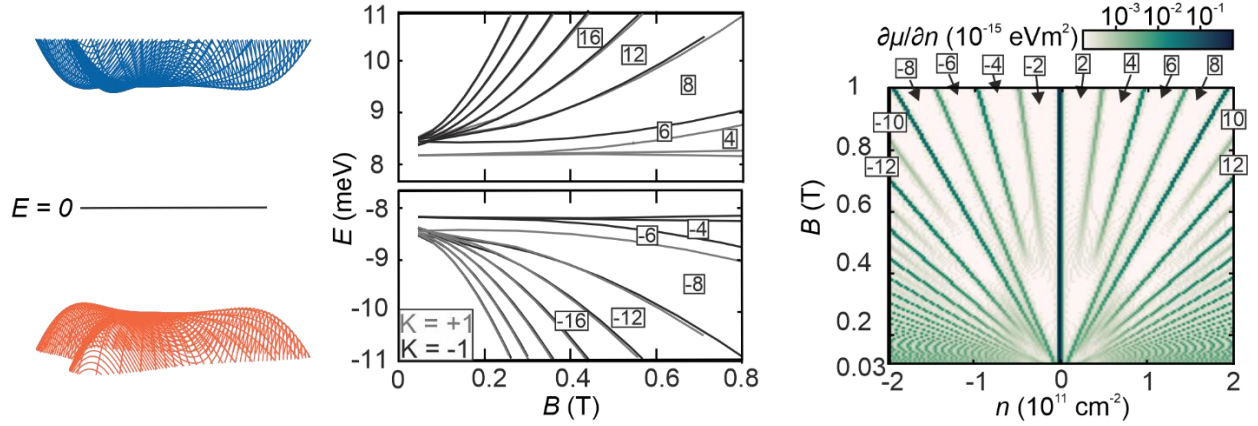
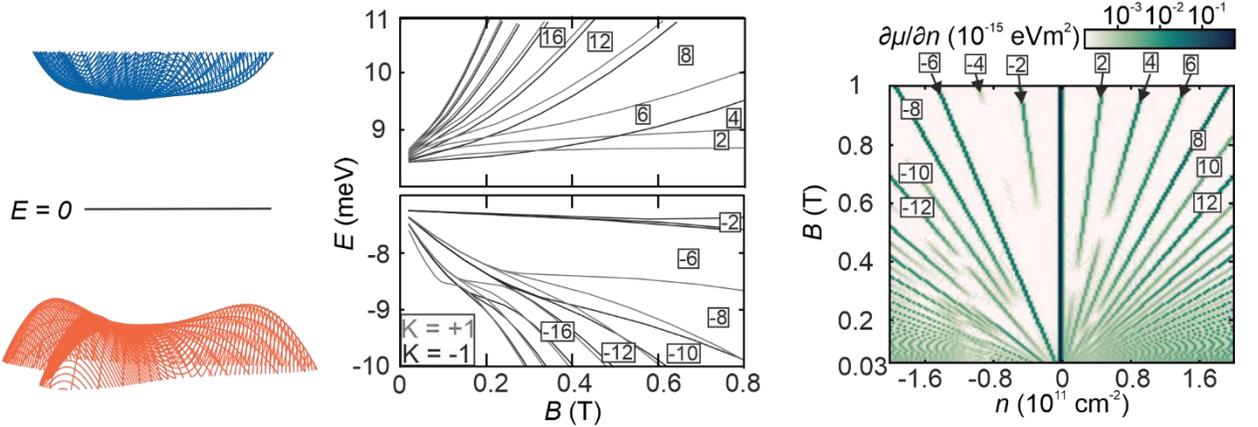

**FIG. S3**. Impact of Coupling parameters $\gamma_3$, $\gamma_4$, and $\Delta'$ on the band structure shown in an energy range from -12 meV to + 12 meV (left panel), the evolution of Landau levels as a function of the magnetic field (middle panel), and the calculated inverse compressibility as a function of charge carrier density and magnetic field (right panel) at $U$ = 0.017 eV. (a) $\gamma_3$, $\gamma_4$, and $\Delta'$ are not included into the calculations. The almost parabolic band structure results in four-fold degenerate Landau level that split up into two-fold degenerate Landau level with increasing magnetic field. This results in the appearance of quantum Hall states with *v* = -4, -2, 0, +2, +4, …. (b) $\gamma_3$ is included into the calculations, $\gamma_4$, and $\Delta'$ are not included. At low energies, the band structure consists of three pockets resulting in a 12-fold degenerate lowest Landau level at low magnetic field and in the appearance of quantum Hall states with *v* = -6 and +6 at very low magnetic fields. Quantum Hall states with *v* = -4, -2, 0, +2, +4, … appear at B > 0.1 T after the magnetic breakdown has occurred. (c) $\gamma_3$, $\gamma_4$, and $\Delta'$ are all included into the calculations. This case is discussed in detail in the main text.

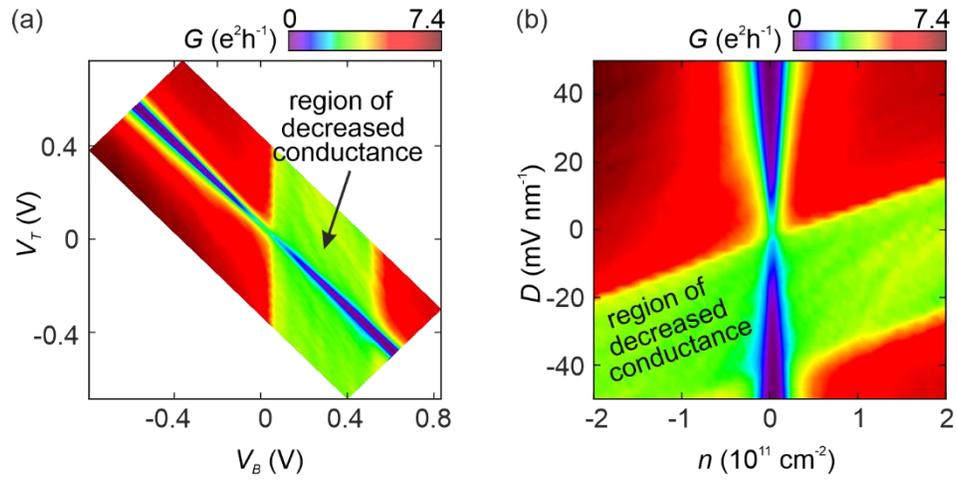

**FIG. S4.** (a) Measured conductance as a function of bottom and top gate. There is a region of decreased conductance that only depends on $V_B$ but not on $V_T$. There is no contact resistance subtracted. b) Measured conductance as a function of density and electric displacement field in the same regime that is shown in (a).

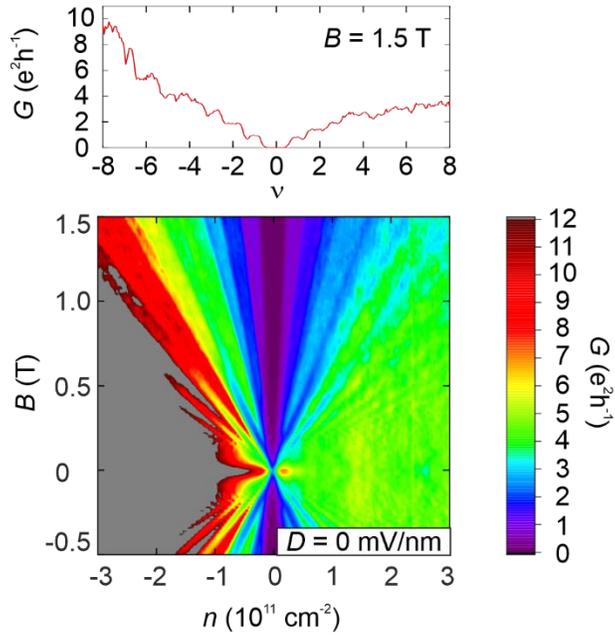

**FIG. S5.** Conductance $G$ as a function of charge carrier density $n$ and magnetic field $B$ in the same $n$- and $B$- space at $D = 0$ mV nm$^{-1}$ as shown in Fig. 1g. A contact resistance of $R_C$ = 3300 Ω - 4600 Ω T · $B^{-1}$ was subtracted from the measured resistance. Using this value of $R_C$ the quantum Hall states with $v < 0$ show a quantized conductance where $G = |v| e^2 h^{-1}$ while the conductance values for $v > 0$ are lower due to a higher contact resistance (see Fig. S4). A line-cut at B = 1.5 T is shown in the top. Here the density $n$ is converted into filling factor $v$.

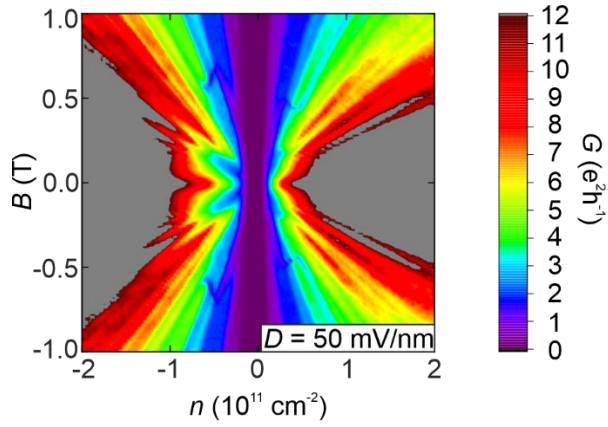

**FIG. S6**. Conductance $G$ as a function of charge carrier density $n$ and magnetic field $B$ at $D$ = 50mV/nm in the same $n$- and $B$- space as shown in Fig. 2e. A contact resistance of $R_C$ = 3000 Ω - 3500 Ω T / $B$ was subtracted from the measured resistance. Using this value of $R_C$ the quantum Hall states with $v$ < 0 show a quantized conductance where $G$ = |$v$| $e^2/h$ while the conductance values for $v$ > 0 are lower due to a higher contact resistance (see Fig. S4).

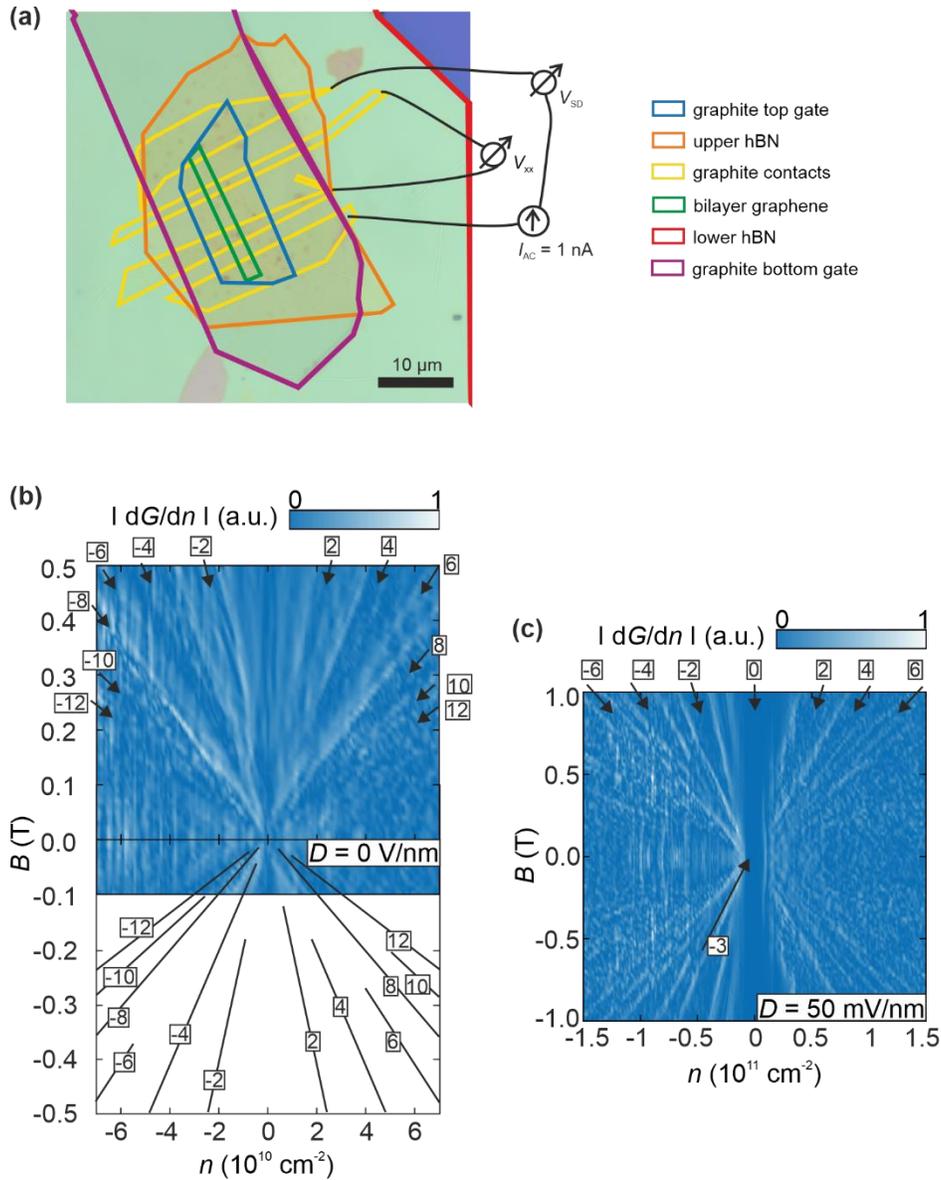

**FIG. S7.** Electrical Measurements conducted in a second device. (a) Optical microscope image of the second device and corresponding configuration of transport measurements. The different flakes are outlined. (b, c) Normalized derivative of the conductance as a function of $n$ and $B$ for $D = 0$ V/nm (b) and $D = 50$ mV/nm (c). Quantum Hall states with $\nu = \pm 8$ and $\nu = -4$ are the most prominent at $D = 0$ V/nm. At $D = 50$ mV/nm quantum Hall states with $\nu = -6$ and $\nu = -3$ are the most prominent for hole doping while the quantum Hall state with $\nu = +4$ is the most prominent for electron doping.